\documentclass{elsart}

\usepackage{color,graphicx}
\usepackage{pstricks,pst-node}
\usepackage{subfigure}
\usepackage{amsmath}
\usepackage{amssymb}
\usepackage{alltt}
\usepackage{setspace}
\usepackage{bbm}

\newcommand{\bra}[1]{\left\langle #1\right|}
\newcommand{\ket}[1]{\left| #1\right\rangle}

\definecolor{Red}{rgb}{1,0,0}
\definecolor{Blue}{rgb}{0,0,1}

\begin{document}
\begin{frontmatter}
\title{Type-II Quantum Algorithms}
\author[addr1]{Peter J. Love\corauthref{cor1}}
\ead{peter@dwavesys.com}
\author{and}
\author[addr1]{Bruce M. Boghosian}
\ead{bruce.boghosian@tufts.edu}
\address[addr1]{Department of Mathematics, Tufts University,\\ Medford, Massachusetts 02155 USA}
\corauth[cor1]{Current Affiliation: D-Wave Systems Inc., Suite 320, 1985 West Broadway, Vancouver, British Columbia, V6J 4Y3 Canada}

\begin{abstract}
We review and analyze the hybrid quantum-classical NMR computing methodology referred to as Type-II quantum computing. We show that all such algorithms considered so far within this paradigm are equivalent to some classical lattice-Boltzmann scheme. We derive a sufficient and necessary constraint on the unitary operator representing the quantum mechanical part of the computation which ensures that the model reproduces the Boltzmann approximation of a lattice-gas model satisfying semi-detailed balance. Models which do not satisfy this constraint represent new lattice-Boltzmann schemes which cannot be formulated as the average over some underlying lattice gas. We close the paper with some discussion of the strengths, weaknesses and possible future direction of Type-II quantum computing. \end{abstract}

\end{frontmatter}

\maketitle

\section{Introduction}

The existence of quantum computers with thousands of error resistant entangled qubits is many years, if not many decades, in the future. However, the success of liquid state nuclear magnetic resonance spectroscopy (NMR) quantum computation means that few ($\sim 10$) entangled qubit quantum computers are available now~\cite{bib:NMR1,bib:NMR2}. This approach exploits the fact that the spin active nuclei in the molecules of a liquid are decoupled from each other and so each nuclear spin may act as a separate quantum bit, and each molecule as a separate few-qubit quantum computer. The states of these nuclear spins may be addressed by radio-frequency pulses. The existence of Avagadro's number of such spins enables the ensemble-averaged results of quantum computations to be obtained via thermodynamic averaging.   

\begin{figure}[htp]
\centering
\begin{pspicture}(0,0)(9,7)

\rput(.5,5.4){\rnode{A1}{}}
\rput(.5,3.4){\rnode{B1}{}}
\rput(.5,1.4){\rnode{C1}{}}
\rput(.75,5.4){\rnode{J}{/}}
\rput(.75,3.4){\rnode{J}{/}}
\rput(.75,1.4){\rnode{J}{/}}

\rput(1,2.25){\rnode{UL1}{}}
\rput(1,.55){\rnode{DL1}{}}
\rput(1.5,2.25){\rnode{UR1}{}}
\rput(1.5,.55){\rnode{DR1}{}}
\rput(1.25,1.5){\rnode{M}{{\Huge I}}}
\rput(1,1.4){\rnode{IC1}{}}
\rput(1.5,2){\rnode{IO3}{}}
\rput(1.5,1.4){\rnode{IO2}{}}
\rput(1.5,0.8){\rnode{IO1}{}}
\ncline{DL1}{UL1}
\ncline{UL1}{UR1}
\ncline{UR1}{DR1}
\ncline{DR1}{DL1}

\rput(1,4.25){\rnode{UL2}{}}
\rput(1,2.55){\rnode{DL2}{}}
\rput(1.5,4.25){\rnode{UR2}{}}
\rput(1.5,2.55){\rnode{DR2}{}}
\rput(1.25,3.5){\rnode{M1}{\Huge I}}
\rput(1,3.4){\rnode{IB1}{}}
\rput(1.5,2.8){\rnode{IO4}{}}
\rput(1.5,3.4){\rnode{IO5}{}}
\rput(1.5,4){\rnode{IO6}{}}
\ncline{DL2}{UL2}
\ncline{UL2}{UR2}
\ncline{UR2}{DR2}
\ncline{DR2}{DL2}

\rput(1,6.25){\rnode{UL3}{}}
\rput(1,4.55){\rnode{DL3}{}}
\rput(1.5,6.25){\rnode{UR3}{}}
\rput(1.5,4.55){\rnode{DR3}{}}
\rput(1.25,5.5){\rnode{M1}{\Huge I}}
\rput(1,5.4){\rnode{IA1}{}}
\rput(1.5,4.8){\rnode{IO7}{}}
\rput(1.5,5.4){\rnode{IO8}{}}
\rput(1.5,6){\rnode{IO9}{}}
\ncline{DL3}{UL3}
\ncline{UL3}{UR3}
\ncline{UR3}{DR3}
\ncline{DR3}{DL3}

\rput(1.75,6){\rnode{J}{/}}
\rput(1.75,5.4){\rnode{J}{/}}
\rput(1.75,4.8){\rnode{J}{/}}
\rput(1.75,4){\rnode{J}{/}}
\rput(1.75,3.4){\rnode{J}{/}}
\rput(1.75,2.8){\rnode{J}{/}}
\rput(1.75,2){\rnode{J}{/}}
\rput(1.75,1.4){\rnode{J}{/}}
\rput(1.75,0.8){\rnode{J}{/}}

\rput(2.25,2){\rnode{Q3}{\psframebox{$Q$}}}
\rput(2.25,1.4){\rnode{Q2}{\psframebox{$Q$}}}
\rput(2.25,0.8){\rnode{Q1}{\psframebox{$Q$}}}

\rput(2.25,4){\rnode{Q6}{\psframebox{$Q$}}}
\rput(2.25,3.4){\rnode{Q5}{\psframebox{$Q$}}}
\rput(2.25,2.8){\rnode{Q4}{\psframebox{$Q$}}}

\rput(2.25,6){\rnode{Q9}{\psframebox{$Q$}}}
\rput(2.25,5.4){\rnode{Q8}{\psframebox{$Q$}}}
\rput(2.25,4.8){\rnode{Q7}{\psframebox{$Q$}}}

\rput(2.75,6){\rnode{J}{/}}
\rput(2.75,5.4){\rnode{J}{/}}
\rput(2.75,4.8){\rnode{J}{/}}
\rput(2.75,4){\rnode{J}{/}}
\rput(2.75,3.4){\rnode{J}{/}}
\rput(2.75,2.8){\rnode{J}{/}}
\rput(2.75,2){\rnode{J}{/}}
\rput(2.75,1.4){\rnode{J}{/}}
\rput(2.75,0.8){\rnode{J}{/}}

\rput(3.25,2){\rnode{M3}{\psframebox{$M$}}}
\rput(3.25,1.4){\rnode{M2}{\psframebox{$M$}}}
\rput(3.25,0.8){\rnode{M1}{\psframebox{$M$}}}

\rput(3.25,4){\rnode{M6}{\psframebox{$M$}}}
\rput(3.25,3.4){\rnode{M5}{\psframebox{$M$}}}
\rput(3.25,2.8){\rnode{M4}{\psframebox{$M$}}}

\rput(3.25,6){\rnode{M9}{\psframebox{$M$}}}
\rput(3.25,5.4){\rnode{M8}{\psframebox{$M$}}}
\rput(3.25,4.8){\rnode{M7}{\psframebox{$M$}}}

\rput(3.75,6){\rnode{J}{/}}
\rput(3.75,5.4){\rnode{J}{/}}
\rput(3.75,4.8){\rnode{J}{/}}
\rput(3.75,4){\rnode{J}{/}}
\rput(3.75,3.4){\rnode{J}{/}}
\rput(3.75,2.8){\rnode{J}{/}}
\rput(3.75,2){\rnode{J}{/}}
\rput(3.75,1.4){\rnode{J}{/}}
\rput(3.75,0.8){\rnode{J}{/}}

\rput(4,2.25){\rnode{UL1}{}}
\rput(4,.55){\rnode{DL1}{}}
\rput(4.75,2.25){\rnode{UR1}{}}
\rput(4.75,.55){\rnode{DR1}{}}
\rput(4.375,1.5){\rnode{blank}{{\Huge A}}}
\rput(4,2){\rnode{AI3}{}}
\rput(4,1.4){\rnode{AI2}{}}
\rput(4,0.8){\rnode{AI1}{}}
\rput(4.75,1.4){\rnode{OC4}{}}
\ncline{DL1}{UL1}
\ncline{UL1}{UR1}
\ncline{UR1}{DR1}
\ncline{DR1}{DL1}

\rput(4,4.25){\rnode{UL2}{}}
\rput(4,2.55){\rnode{DL2}{}}
\rput(4.75,4.25){\rnode{UR2}{}}
\rput(4.75,2.55){\rnode{DR2}{}}
\rput(4.375,3.5){\rnode{blank}{\Huge A}}
\rput(4,4){\rnode{AI6}{}}
\rput(4,3.4){\rnode{AI5}{}}
\rput(4,2.8){\rnode{AI4}{}}
\rput(4.75,3.4){\rnode{OB4}{}}
\ncline{DL2}{UL2}
\ncline{UL2}{UR2}
\ncline{UR2}{DR2}
\ncline{DR2}{DL2}

\rput(4,6.25){\rnode{UL3}{}}
\rput(4,4.55){\rnode{DL3}{}}
\rput(4.75,6.25){\rnode{UR3}{}}
\rput(4.75,4.55){\rnode{DR3}{}}
\rput(4.375,5.5){\rnode{blank}{\Huge A}}
\rput(4,6){\rnode{AI9}{}}
\rput(4,5.4){\rnode{AI8}{}}
\rput(4,4.8){\rnode{AI7}{}}
\rput(4.75,5.4){\rnode{OA4}{}}
\ncline{DL3}{UL3}
\ncline{UL3}{UR3}
\ncline{UR3}{DR3}
\ncline{DR3}{DL3}

\rput(5.5,5.4){\rnode{J}{/}}
\rput(5.5,3.4){\rnode{J}{/}}
\rput(5.5,1.4){\rnode{J}{/}}


\rput(6,6.25){\rnode{UL3}{}}
\rput(6,.75){\rnode{DL3}{}}
\rput(7.5,6.25){\rnode{UR3}{}}
\rput(7.5,.75){\rnode{DR3}{}}
\rput(6.75,3.5){\rnode{blank}{\Huge C}}
\rput(6,5.4){\rnode{IA5}{}}
\rput(7.5,5.4){\rnode{OA5}{}}
\rput(6,3.4){\rnode{IB5}{}}
\rput(7.5,3.4){\rnode{OB5}{}}
\rput(6,1.4){\rnode{IC5}{}}
\rput(7.5,1.4){\rnode{OC5}{}}
\ncline{DL3}{UL3}
\ncline{UL3}{UR3}
\ncline{UR3}{DR3}
\ncline{DR3}{DL3}

\rput(7.75,5.4){\rnode{J}{/}}
\rput(7.75,3.4){\rnode{J}{/}}
\rput(7.75,1.4){\rnode{J}{/}}


\rput(8,5.4){\rnode{OAF}{}}
\rput(8,3.4){\rnode{OBF}{}}
\rput(8,1.4){\rnode{OCF}{}}

\ncline[linewidth=2pt]{A1}{IA1}
\ncline[linewidth=2pt]{B1}{IB1}
\ncline[linewidth=2pt]{C1}{IC1}
\ncline{OA1}{IA2}
\ncline{OB1}{IB2}
\ncline{OC1}{IC2}
\ncline{OA2}{IA3}
\ncline{OB2}{IB3}
\ncline{OC2}{IC3}
\ncline{IO1}{Q1}
\ncline{IO2}{Q2}
\ncline{IO3}{Q3}
\ncline{IO4}{Q4}
\ncline{IO5}{Q5}
\ncline{IO6}{Q6}
\ncline{IO7}{Q7}
\ncline{IO8}{Q8}
\ncline{IO9}{Q9}
\ncline{Q1}{M1}
\ncline{Q2}{M2}
\ncline{Q3}{M3}
\ncline{Q4}{M4}
\ncline{Q5}{M5}
\ncline{Q6}{M6}
\ncline{Q7}{M7}
\ncline{Q8}{M8}
\ncline{Q9}{M9}
\ncline{M1}{AI1}
\ncline{M2}{AI2}
\ncline{M3}{AI3}
\ncline{M4}{AI4}
\ncline{M5}{AI5}
\ncline{M6}{AI6}
\ncline{M7}{AI7}
\ncline{M8}{AI8}
\ncline{M9}{AI9}

\ncline[linewidth=2pt]{OA4}{IA5}
\ncline[linewidth=2pt]{OB4}{IB5}
\ncline[linewidth=2pt]{OC4}{IC5}
\ncline[linewidth=2pt]{OA5}{OAF}
\ncline[linewidth=2pt]{OB5}{OBF}
\ncline[linewidth=2pt]{OC5}{OCF}
\end{pspicture}
\caption{A single step of a Type-II quantum lattice-gas computation. Thick wires indicate classical registers. Thin wires indicate quantum registers. The computation proceeds by initializing a set of parallel quantum computations (the boxes ``Q'') representing a superposition over the states of a given site of a classical lattice gas. The collision step is performed by a set of independent ensembles of quantum computations, one ensemble for each lattice site. The results of the measurements are a set of single particle occupation probabilities for the lattice vectors at each site, (classical information), are input to a classical computer (the box ``C''), which performs the propagation step of the lattice gas algorithm.\label{fig:nmrt2}}
\end{figure}
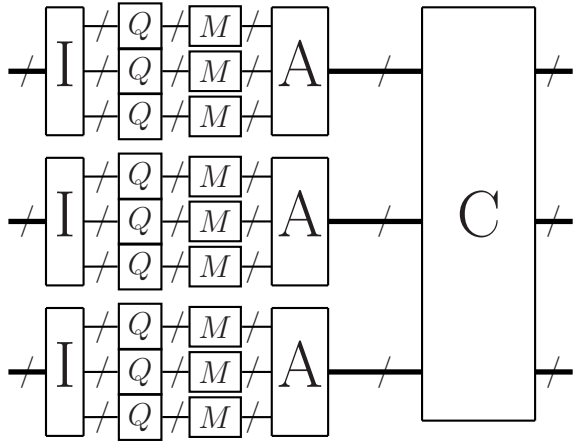

The resonant frequency depends on the local magnetic field, and so placing the fluid in a non-uniform field enables one to separately address different regions of fluid~\cite{bib:yepez6}. Each region of fluid then acts as a separate ensemble of quantum computers, and the idea of using a large array of these computers was first proposed in~\cite{bib:yepez12}. Quantum coherence is maintained within each node on the array, and the arrays are measured after a few quantum coherent  operations and then classical information representing the ensemble average of $\sim 10^{18}$ measurements is exchanged between nodes, providing the initial state for the next coherent quantum computations on each node. This situation is illustrated in Figure~\ref{fig:nmrt2}. 

Type II quantum computers were introduced in the context of physical modelling, and all subsequent work has been focussed towards implementation of some lattice based algorithm for (usually classical) physics~\cite{bib:yepez4}.  Algorithms have been proposed for fluid dynamics~\cite{bib:yepez10,bib:yepez11,bib:yepez12}, for the diffusion equation~\cite{bib:yepez9}, the Burgers equation~\cite{bib:yepez7}, for the nonlinear Schroedinger and Korteweg-de Vries equations in one-dimensions~\cite{bib:yepez1}, and for MHD turbulence in one dimension~\cite{bib:yepez2}. Some of these algorithms have also been implemented on NMR quantum computers~\cite{bib:yepez3,bib:yepez6,bib:chenburgers}.

\section{Type II Quantum Lattice Gases}\label{sec:TII}

The lattice structure of the array of quantum processors and the similarity of the two step quantum computation/classical communication are very similar to the collision/propagation evolution of lattice-gas and lattice-Boltzmann algorithms~\cite{bib:doolenreview,bib:saurobook}. Because of this similarity Type-II algorithms are described in the literature as quantum lattice-gases. One must distinguish these Type-II models from the quantum lattice gases for the Dirac and Schroedinger equation, designed to realize exponential improvements in performance on fully coherent quantum computers~\cite{bib:meyer2,bib:bogwash3}. Type II quantum lattice-gases perform the collision step using quantum computation, but then sample the results of those computations before performing the subsequent propagation. Type II simulations rely on ensemble averaging, and these simulations exploit the natural existence of an ensemble within the NMR quantum computer to estimate the average values of the qubits at a site. Note carefully, however, that only information about classical averages is communicated between sites. Information about two-particle and higher correlations is not preserved by this measurement and communication step, and so the results of a Type II quantum lattice gas simulation are equivalent to the results of a classical lattice Boltzmann simulation. Indeed, they most closely resemble the early lattice-Boltzmann algorithms which used ensemble averaged lattice gas collision rules at each site~\cite{bib:higuera}.

All of the models implemented so far on NMR quantum computers have two qubits per site in one dimension. The state of the model at a site is given by a pair of single particle distribution functions, $(f_1,f_2)$, such that the probability that bit one is one is $f_1$ and the probability that bit two is one is $f_2$. Each separate quantum processor therefore consists of a pair of coherent qubits. The state of these qubits is initialized using $(f_1,f_2)$ as follows:
\begin{equation}
\ket{\psi(t=0)} = \sqrt{(1-f_1)(1-f_2)}\ket{00}+\sqrt{(1-f_1)f_2}\ket{01}+\sqrt{(f_1(1-f_2)}\ket{10}+\sqrt{f_1f_2}\ket{11}
\label{eq:purestate}
\end{equation}

The probabilities of the four possible lattice gas states at a site, $\{00,01,10,11\}$ are ${\bf n}=((1-f_1)(1-f_2),(1-f_2)f_1,f_2(1-f_1)  ,f_1f_2)$ in the Boltzmann approximation. The diagonal components of the density matrix $\ket{\psi(t=0)}\bra{\psi(t=0)}$ are equal to these probabilities. The collision step of a stochastic lattice gas model is given by the multiplication of the vector {\bf n} by a matrix, A, satisfying normalization and semi-detailed balance (a doubly stochastic or complete matrix). In our Type-II simulation we wish to reproduce the effect of this matrix multiplication by conjugating the density matrix by the unitary operator corresponding to the quantum mechanical part of our Type-II algorithm. The effect of this conjugation on the diagonal elements of the density matrix must be the same as the effect of matrix multiplication of vector ${\bf n}$ by the matrix $A$. 

Now consider a Type-II quantum computer with $b$ qubits per node. At the beginning of each simulation step the qubits are initialized with a density matrix whose entries correspond to the probabilities of the qubit state in the molecular chaos approximation. After each local unitary update, the qubits are measured and because the existence of an ensemble is presumed in all Type-II algorithms the probability that qubit $i$ is one is associated with the single particle distribution function $f_i({\bf x})$ at site ${\bf x}$ and vector $i$. During the collision step the unitary action $U$  is applied to this density matrix and it evolves to:
\begin{equation}\label{eq:typeIIcoll}
\rho_{mr} \mapsto \sum_{mr}U_{pm}\rho_{mr} \left[U^\dagger\right]_{rq}
\end{equation}

The density matrix formulation of quantum mechanics contains within it classical states, real linear convex combinations of classical states representing classical probability distributions, pure quantum states and mixed quantum states. For this reason Type-II algorithms may be constructed which reproduce deterministic reversible lattice gas rules, stochastic lattice gas rules, and new models which have collision operators which do not correspond to the classical average over some lattice gas. In all cases, because the measurement and averaging step precedes the classical communication step, the overall algorithm is equivalent to a classical lattice-Boltzmann scheme with collision operator given by~(\ref{eq:typeIIcoll}).

Taking our update for the density matrix and restricting it to an update of the diagonal components of the density matrix:
\begin{equation}
\rho'_{pp}= \sum_{mr}U_{pm}\rho_{mr} \left[U^\dagger\right]_{rp}
\end{equation}
Where there is no implicit sum on repeated indices. Separating the action on the diagonal and off diagonal components of the density matrix:
\begin{equation}
\rho'_{pp}= \sum_{m}U_{pm}U^*_{pm}\rho_{mm} +\sum_{m\neq r}U_{pm}U^*_{pr}\rho_{mr} 
\end{equation}

If the second term here is zero then the action of the unitary operator $U$ on the diagonal components of $\rho$ is identical to the action of a matrix $A$ whose components are the modulus squared of the components of $U$.
\begin{equation}
A_{pm} = U_{pm}U^*_{pm}
\end{equation}
The unitarity of $U$ implies that the matrix $A$ will obey normalization and semi-detailed balance. 

It remains to formulate a sufficient constraint that the second term above is zero. A sufficient condition for the Type-II algorithm to reproduce the action of a stochastic lattice-gas collision operator whose average is the matrix $A$ is that:
\begin{equation}
\mbox{Re}
[{U_{pm}U^*_{pr}}]=0\hspace{0.5cm} \forall m<r, \forall p
\label{eq:TIIconstraint}
\end{equation}

One realization of a Type-II algorithm on an NMR quantum computer implemented a simple stochastic lattice-gas algorithm for the diffusion equation in one-dimension, where the matrix $A$ is:
\begin{equation}
A=\frac{1}{2}\begin{pmatrix}2&0&0&0\\0&1&1&0\\0&1&1&0\\0&0&0&2\end{pmatrix}
\end{equation}
In~\cite{bib:yepez6,bib:yepez3} this model is implemented using a unitary matrix $U$:
\begin{equation}
U=\begin{pmatrix}1&0&0&0\\0&\frac{1-i}{2}&\frac{1+i}{2}&0\\0&\frac{1+i}{2}&\frac{1-i}{2}&0\\0&0&0&1\end{pmatrix}
\end{equation}
It is clear by inspection that the modulus squared of the components of $U$ give the components of $A$. The condition~(\ref{eq:TIIconstraint}) is also satisfied.

If a Type-II algorithm does not obey constraint~(\ref{eq:TIIconstraint}) the action of $U$ on the diagonal components of the density matrix may not be described by a doubly stochastic matrix. The new values of the diagonal components will be modified by terms involving the off diagonal components of the density matrix, which in turn will involve the square roots of the single particle distribution functions. The action on the diagonal components of an arbitrary density matrix will not be linear and so~(\ref{eq:TIIconstraint}) is a necessary as well as sufficient condition. Models violating~(\ref{eq:TIIconstraint}) fall outside the class of lattice Boltzmann models which may be considered as averages over some underlying lattice-gas model in the Boltzmann approximation, and a detailed analysis of such lattice-Boltzmann models remains an open problem.  An example of such a model is the lattice-gas model for the Burgers equation~\cite{bib:bogtaylor}, which does not satisfy semi-detailed balance. A Type-II algorithm has been proposed for this model, and implemented in NMR~\cite{bib:yepez7,bib:chenburgers}. 

In the case where the ensemble of molecular quantum computers at each node is initialized, not in the pure state~(\ref{eq:purestate}), but in the completely mixed state with a density matrix with diagonal components equal to ${\bf n}$ and off diagonal components which are all zero, any unitary action reproduces the collision operator of a stochastic lattice gas satisfying detailed balance and normalization. In this case the Type-II computation is reduced to simple molecular computing of a Boltzmann approximation to a lattice-gas evolution, and no novel models are possible.

\section{Conclusions}

The characterization of current Type-II simulations as equivalent to classical lattice-Boltzmann methods illustrates the limited utility of Type-II quantum computation as presently defined. These algorithms provide at best a constant speedup and require an ensemble in order to accurately sample the classical information available through measurement which is exchanged between quantum processors. While applying techniques from medical imaging may make the realization of very large ($>2048^3$) grids possible, such techniques require a three dimensional field gradient to be useful. While the implementation of such field gradients is non-trivial, such large grid sizes would challenge the largest lattice Boltzmann simulations on classical supercomputers to date.

One of the strengths of Type II quantum computation is that it takes advantage of the ensemble inherent in NMR implementations. Distribution functions are obtained by averaging over the $\sim 10^{18}$ molecules in each small region~\cite{bib:nmrreview}. Of course, this necessity to obtain the average values by sampling becomes an Achilles heel once one considers more scalable architectures (such as superconducting or solid state optically addressed implementations) where the necessity for sampling requires multiplication of the hardware. Although the spirit of Type II computing is that many small quantum computers are likely to be available sooner than a single large machine, the more scalable architectures are unlikely to reproduce the huge ensemble freely available with liquid state NMR.

From the above analysis we can see that unitary matrices obeying~(\ref{eq:TIIconstraint}) implement a doubly-stochastic Markov matrix action on the diagonal components of the density matrix. The size of this Markov matrix grows exponentially with  the number of qubits. It is natural to ask whether one may use these Type-II techniques to solve Markov chain problems out of the reach of classical computers. For example, if we could address and control a pseudo-pure state of an ensemble of $100$ qubit systems we could address Markov problems of size $2^{100}$, which are out of reach of any conceivable classical digital computer. 
However, the limit on the size of Markov process we may implement is not given by the number of qubits but by our ability to read out the diagonal components of the density matrix by sampling from the ensemble. We require at least one member of the ensemble per state of the system in order to do this, which implies that for $100$ qubits one requires $10^7$ moles of material in our ensemble. Unfortunately this is not feasible. Current liquid state NMR realizations utilize $10^{18}$ molecules, implying a maximum size of problem of $2^{60}$, implying that the maximum Markov problem size occurs for such an ensemble with $60$ qubits. Such problems are classically intractable above at most $2^{30}$, implying that there is a range of problems which can be implemented on small NMR ensemble quantum computers which can coherently address $30-60$ spins which are classically intractable for conventional digital computers.

This observation reinforces the point that Type-II quantum computation is really an interesting mixture of quantum computing with classical molecular computing. The approach benefits as much from the presence of a large number of realizations of the computation in a molecular ensemble as it does from the exponential growth of the dimension of the Hilbert space of the individual quantum processors.

This material is based on work supported in part by the U.S. Army Research Laboratory and the U.S. Army Research Office under grant number W911NF-04-1-0334, in part by the U.S. Air Force Office of Scientific Research under grant number FA9550-04-1-0176, and in part by the DARPA QuIST Program under AFOSR grant number F49620-01-1-0566.

\end{document}